\documentclass[letterpaper, 12pt]{article}

\usepackage{cmjStyle}
\usepackage{natbib}
\bibpunct{(}{)}{;}{a}{}{, }
\raggedright
\usepackage{epstopdf}
\usepackage{amsmath,amssymb,amsbsy,bm,upgreek,nicefrac}
\usepackage{todonotes,microtype}

\providecommand{\tightlist}{%
  \setlength{\itemsep}{0pt}\setlength{\parskip}{0pt}}
\usepackage{longtable,booktabs,array,calc,etoolbox}
\makeatletter
\patchcmd\longtable{\par}{\if@noskipsec\mbox{}\fi\par}{}{}
\newsavebox\pandoc@box
\newcommand*\pandocbounded[1]{%
  \sbox\pandoc@box{#1}%
  \Gscale@div\@tempa{\textheight}{\dimexpr\ht\pandoc@box+\dp\pandoc@box\relax}%
  \Gscale@div\@tempb{\linewidth}{\wd\pandoc@box}%
  \ifdim\@tempb\p@<\@tempa\p@\let\@tempa\@tempb\fi
  \ifdim\@tempa\p@<\p@\scalebox{\@tempa}{\usebox\pandoc@box}%
  \else\usebox{\pandoc@box}%
  \fi}
\def\fps@figure{htbp}
\makeatother

\graphicspath{{./figures/}}
\makeatletter
\renewcommand\hyper@natlinkbreak[2]{#1}
\makeatother
\usepackage{placeins}
\begin{document}

{\cmjTitle How Close to the Theme? Similarity-Label Conditioning for
Game Music Generation}
\vspace*{24pt}

{\cmjAuthor Shangxuan Luo, Zixun Guo, Jinjie Shi, Joshua Reiss\par}
\begin{cmjAuthorAddress}
Centre for Digital Music, Queen Mary University of London, UK \\
s.luo@qmul.ac.uk
\end{cmjAuthorAddress}

\begin{abstract}
Composers develop music from an original idea, controlling how similar
each new section is to it, from near repetition to new material. Game
music needs this often, as its material repeats over long play. Music
models, however, rarely learn to control similarity from real pieces. We
further train a pretrained MIDI model on section pairs from Super Mario
pieces. A label states each pair's similarity, and a context vector
summarises the given section, or context. For 157 test contexts, the
model generates six outputs per label. Twelve measures score the
outputs; one main measure each covers melody, pitch content, rhythm and
note length. On average, outputs requested as similar are more similar
to their context than those requested as dissimilar, and not to other
pieces. All twelve measures favour them; each main measure stays
significant after correction for multiple tests. On melody, the gap
comes mainly from more frequent restatement of the context's melody.
Against an otherwise identical model without a label, only the request
for high similarity makes outputs more similar. In a listening test,
ratings followed melodic overlap with the context most closely of all
measures.
\end{abstract}

\section{Introduction}\label{introduction}

Music is built by repeating and varying its own material, and the
variation lets a listener hear two passages as related
\citep{volk2012modelling}. The composer controls how similar each new
section is to that material, from near repetition to a new idea.

Game music asks for this choice more often than most music. It is
assembled from loops, branches and layers that the game recombines
during play \citep{summers2016understanding}. Composers write
interchangeable sections over the same material, so the game can switch
between them \citep{winifred2017composers}. Too exact a repetition
becomes noticeable over long sessions
\citep{lopezduarte2024progressiveadaptive, worrallExpressivePerformanceRendering2025}.
A tool should therefore let the composer set this similarity, since
composers want creative control over AI tools
\citep{worrall2023considerations}.

Symbolic music models continue a passage fluently but do not offer this
choice. Systems for variation let a user set how strongly an original is
altered \citep{bhandariImprovNetGeneratingControllable2025}. Real
pieces, however, show how close their own sections stand to one another.
One model learns variation from such pieces, but only from pairs within
fixed similarity bounds, without a choice of degree
\citep{gaoVariationTransformerNew2024}. To our knowledge, no model
learns to set similarity from real pairs across their range.

This paper asks whether a model can learn this control from section
pairs labelled by their similarity. We fine-tune, or further train, the
Moonbeam MIDI model on pairs from one Super Mario piece: a given
section, the context, and a target. Each pair carries a high-, mid- or
low-similarity label from compound similarity (CS), a measure of our
own, the label measure. A \emph{context vector} summarising the context
enters the model at every bar. Three controls separate their roles: one
lacks the label, one lacks label and context vector, one has wrong
labels. Each model generates six outputs per context and label, one per
sampling run, or draw. Twelve measures of melody, pitch content, rhythm
and note length, nine computed differently from CS, and a listening test
judge the outputs.

Averaged over draws, outputs requested as highly similar are more
similar to their context than those requested as dissimilar. All twelve
measures favour them, and the four main measures and CS remain
significant after correction for multiple tests. Against the no-label
control, only the high-similarity label makes outputs more similar. On
melody, pitch content and rhythm the shift is toward the output's own
context, not other pieces. On melody it comes mainly from more frequent
restatement of the context's melody. Mid-similarity outputs do not lie
in between.

The listening test did not detect the label effect: its 30 pairs, one
output per label, differed little on the measures. Listener ratings
followed melodic overlap most closely, which CS leaves out; only 38
percent of its high-similarity pairs restate the same section.

The paper makes two contributions. The first is a model whose similarity
label is learned from real section pairs. Requesting high similarity
makes its outputs more similar to their own context on average. The
second is evidence from ratings of 10-second clips that melodic overlap
has the largest observed association with judged similarity. It
motivates a melody-aware label measure, which we leave to future work.

\section{Related Work}\label{related-work}

Related work answers three questions: why graded similarity matters in
game music, how far models already control it, and how it can be
measured.

\subsection{Game Music}\label{game-music}

Surveys of generative music for games describe rule-based and learned
systems \citep{plut2020generative, shangxuanluo2025procedural}. Most
condition on the player's or the game's state: an emotion model
\citep{ferreira2020computergenerated, amaral2022adaptive}, the player's
arousal \citep{tanabe2023adaptive}, gameplay variables
\citep{lopezduarte2024progressiveadaptive} or biofeedback
\citep{mcintoshAffectiveConditionalModifiers2023}. None asks how closely
a new section should resemble one already played.

\subsection{Controllable Generation}\label{controllable-generation}

Symbolic generation has moved toward control. Music FaderNets
\citep{tan2020music} steer arousal through sliders for rhythm and note
density. Pretrained models such as Moonbeam \citep{guo2025moonbeam} and
NotaGen \citep{wang2025notagen} are fine-tuned for narrower tasks. A
common route controls attributes such as style or emotion
\citep{ji2024survey}; FIGARO \citep{rutteFIGAROGeneratingSymbolic2024}
matches bar-level descriptions such as note density and chords.
Structure-conditioned models take a form
\citep{daiControllableDeepMelody2021, wangWholeSongHierarchical2024} or
a theme \citep{shihThemeTransformerSymbolic2023} as input. They decide
whether material repeats, not how close a new section is to an old one.

Several approaches address closeness. MusicVAE
\citep{roberts2018hierarchical} interpolates in a latent space, and
ImprovNet \citep{bhandariImprovNetGeneratingControllable2025} sets how
far an improvisation departs by the strength of a corruption step.
MuseMorphose \citep{wuMuseMorphoseFullSong2021} alters a given piece
through bar-level attributes such as rhythmic intensity. These define
closeness by a model mechanism or by attributes. Infilling models such
as the Anticipatory Music Transformer
\citep{thickstunAnticipatoryMusicTransformer2023} regenerate a span
around fixed events, without a control of its similarity to the
original. Variation Transformer \citep{gaoVariationTransformerNew2024}
learns from theme-and-variation pairs within real pieces, game music
among them, but only within a fixed similarity range. Our labels span
the training pairs' range and relate two sections, unlike FIGARO's and
FaderNets' attributes, which describe one passage.

\subsection{Similarity Measurement}\label{similarity-measurement}

No approach to musical similarity serves all music
\citep{volk2012modelling, lerch2026survey}; game music includes ambient
tracks \citep{winifred2017composers} and atonal cues
\citep{summers2016understanding}. Melodic matching compares n-grams,
runs of n successive intervals, and edit distances
\citep{uitdenbogerdMelodicMatchingTechniques1999}, and weighted
combinations of such measures can approximate expert judgements
\citep{mullensiefenOptimizingMeasuresMelodic2004}. Listeners abstract
cues, salient elements at the musical surface
\citep{deliegeSimilarityPerceptionCategorization2001}. Their similarity
ratings of piano extracts followed surface features such as dynamics,
texture and contour more than motivic or harmonic relations
\citep{lamontMotivicStructurePerception2001}. For folk melodies,
distributional pitch features explained 40\% of the ratings, and with
melodic predictability, 55\%
\citep{eerolaStatisticalFeaturesPerceived2001}. Listening tests remain
the usual evaluation, without a standard procedure
\citep{lerch2026survey}. This paper therefore labels pairs with one
measure and judges outputs with others and listeners.

\section{Data}\label{data}

The data are Super Mario piano scores, sectioned by an AI and checked by
two annotators.

\subsection{Corpus}\label{corpus}

The corpus holds the 554 Super Mario pieces on NinSheetMusic
(www.ninsheetmusic.org), an archive of piano transcriptions. We excluded
39 pieces written for two pianos or for mixed instruments. Of the
remaining 515, 334 have two or more main-body sections, which yield
pairs. They are split into 236 training, 49 validation and 49 test
pieces, stratified by section count. Arrangements of one tune,
identified by title and length, stay in one set.

\subsection{Section Annotation}\label{section-annotation}

Gemini 3.1 Pro, a model of the Gemini family
\citep{teamGeminiFamilyHighly2025}, annotated every piece, reading its
score as ABC text. It divides a piece into sections and gives each a
function, such as intro or stinger \citep{summers2016understanding} or
main body, and a letter. Within the main body, sections that share their
melodic core and rhythm take the same letter.

\subsection{Human Validation}\label{human-validation}

Two annotators labelled 47 pieces, drawn at random across piece lengths,
independently and without seeing the AI annotation. They marked nested
levels of segments, where a shared letter within a level marks the same
material. An AI section is consistent with an annotator when it equals
one human segment, or a run of consecutive segments, at some level. Two
AI sections are the same material for an annotator when their letters
agree. Figure S1 in the supplement shows an example.

The AI agrees moderately to substantially with each annotator, on the
usual scale \citep{landisMeasurementObserverAgreement1977}, about as
well as the annotators agree. Only main-body sections are compared,
since the schemes label introductions, endings and silence differently.
Both annotators judge 128 such pairs. Cohen's kappa
\citep{cohenCoefficientAgreementNominal1960}, agreement beyond chance,
is 0.59 between the AI and the first annotator (95 percent interval
0.25--0.81) and 0.68 with the second (0.44--0.84). The annotators reach
0.58 (0.35--0.76). Of the 177 AI sections, 87\% and 82\% are consistent
with the two annotators. The wide intervals fit parity, not superiority.

\section{Similarity Labels}\label{similarity-labels}

The model learns graded similarity from labelled section pairs. Each
pair joins a context and a target section from one piece; its label
states how similar they are.

\subsection{Compound Similarity}\label{compound-similarity}

Since musical similarity has no agreed definition, a labelling function
must choose what it compares. A linear combination of similarity
measures whose weights sum to one is itself a similarity measure. Such
combinations have matched expert judgements of melodic similarity
\citep{mullensiefenOptimizingMeasuresMelodic2004}. We label pairs with
compound similarity (CS), a measure of our own built in this way. CS
assumes that the similarity of two sections can be approximated by four
distributions of their notes.

CS-Chroma compares the two \emph{chromas}, 12-bin histograms of pitch
classes that ignore the octave, with each note weighted by its duration.
Their cosine similarity is taken at the best of twelve transpositions,
so that a theme restated in another key still scores as similar.
CS-Duration compares note lengths by cosine similarity over 16
logarithmic bins. The bins run between each section's own shortest and
longest note, so CS-Duration compares the shapes of the two
distributions, not shared durations. CS-Register compares the mean
pitches: one minus their difference over 48 semitones, floored at zero.
CS-Density compares the notes per unit time: one minus their difference
over the larger rate. Every component but CS-Register is invariant to
transposition; Table S1 gives each formula.

CS is the weighted mean of the components, with relative weights 4, 3, 1
and 1 in this order. The weights were set heuristically when CS was
designed, before any evaluation, and were not tuned. Melody is not a
component; the Discussion gives the reasons.

\subsection{Label Thresholds}\label{label-thresholds}

The 334 pieces yield 3,304 ordered pairs: every two main-body sections
of a piece, in both orders. Two thresholds, 0.78 and 0.60, divide the
scores into three labels of near-equal size (Figure 1). They are the
rounded tertiles of all 3,304 pairs, test pieces included. We write them
\texttt{sim:high}, \texttt{sim:mid} and \texttt{sim:low}, and call them
the high-, mid- and low-similarity labels below. They fix the training
labels and also group the real targets with which the evaluation
compares the outputs.

\begin{figure}
\centering
\includegraphics[width=0.55\linewidth,height=\textheight,keepaspectratio,alt={Most
pairs are moderately similar, and the low and middle labels split one
dense peak. The distribution of CS over all 3,304 pairs; dashed lines
mark the two thresholds. The high-similarity label holds 1,090 pairs,
the middle one 1,172 and the low one 1,042.}]{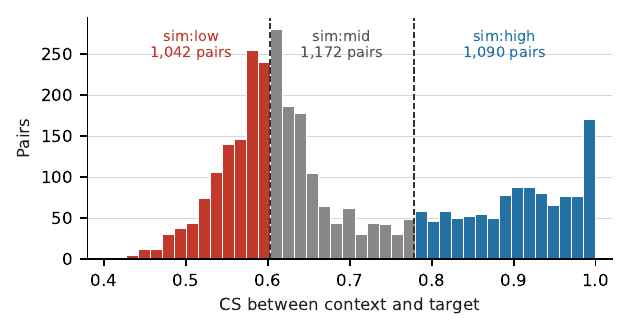}
\caption{Most pairs are moderately similar, and the low and
middle labels split one dense peak. The distribution of CS over all
3,304 pairs; dashed lines mark the two thresholds. The high-similarity
label holds 1,090 pairs, the middle one 1,172 and the low one 1,042.}
\end{figure}

\subsection{Annotation Check}\label{annotation-check}

The section annotation offers an independent check. Two sections that
share a letter, such as two A sections, are versions of the same
material. A similarity measure should score such pairs (AA) above
different-letter pairs (AB). CS passes this check. Over all 3,304 pairs,
a random AA pair outscores a random AB pair with probability 0.834 (95
percent interval 0.80--0.87, resampling whole pieces). The labels also
order the annotation. Same-letter pairs make up 38.2\% of the
high-similarity label, 17.6\% of the middle one and 1.2\% of the low
one. Nor does it depend on the AI letters. Both annotators can place 254
ordered pairs of 30 pieces. On them, CS scores 0.671 and 0.710 under
their letters and 0.741 under the AI letters. The subset, more than the
letters, lowers these values.

\section{Model}\label{model}

SimLlama adapts a pretrained symbolic music model to continue from a
context section toward a requested similarity. A label states the
request, and a context vector summarises the context; Figure S2 in the
supplement draws the whole input.

\subsection{Base Model}\label{base-model}

The base model is Moonbeam 839M \citep{guo2025moonbeam}, a MIDI
foundation model pretrained on about 18 billion tokens. Its backbone is
the LLaMA transformer \citep{touvronLLaMAOpenEfficient2023} with
Moonbeam's own relative attention. SimLlama takes its name from this
backbone; its weights come from Moonbeam. A small recurrent decoder
emits each note as six fields: onset, duration, octave, pitch class,
instrument and velocity.

\subsection{Conditioning and Training}\label{conditioning-and-training}

The input places the context notes first, then a separator and one of
the three label tokens, then the target notes. The loss covers only the
target notes. We expected the label alone to give the model only a weak
hold on the context, which lies hundreds of tokens back. A context
vector supplies a direct one. The backbone encodes the context, and the
mean of its last hidden layer, mapped by a learned linear projection, is
the context vector. At each bar boundary of the target, a CTX token
takes this vector as its embedding.

The backbone stays frozen. Low-rank adapters
\citep{huLoRALowRankAdaptation2021a}, small trainable matrices added to
a frozen network, adapt its query and value projections. With the
projection and the new token embeddings, 4.60 million parameters train
(Table S2). Four models train from one checkpoint on the same pairs and
see the context notes first; a fifth is in the supplement. SimLlama
receives the correct label and the context vector. The no-label control
drops the label, so it differs from SimLlama in the label and its
stopping epoch alone. The wrong-label control keeps both signals, but in
training its label is always one of the two wrong ones, fixed per pair.
The unconditioned control receives neither; against the no-label
control, it isolates the context vector. Early stopping keeps the
checkpoint with the lowest validation loss (Table S2). SimLlama reached
it at epoch 11.

\subsection{Generation}\label{generation}

Each draw generates one section per context and label. The note budget
is 1.2 times the mean length of that context's test targets in notes. It
caps the length, the same for every label; a composer would set it by
hand. Each of a note's six fields is sampled at temperature 0.8 with
nucleus sampling at p = 0.9, which keeps the most probable choices.

\section{Objective Evaluation}\label{objective-evaluation}

The evaluation asks whether outputs requested as highly similar are more
similar to their context than the others. As no similarity measure
serves all music, we use several, one per musical dimension. Each comes
from a published method and was checked against the annotations and,
after the fact, the ratings.

\subsection{Design}\label{design}

The design fixes the outputs, the measures that score them and the
tests.

\subsubsection{Test Material}\label{test-material}

For each of the 157 test contexts, SimLlama generates one section under
each label in six seeded draws. The controls draw six times per label,
or twelve times without one. A draw of a context counts only if all
three of its outputs hold at least ten notes, so that every measure can
be computed. Of the 942 draws, 929 pass, and each context averages its
passing draws.

Post-processing cuts every output at the context's length and repairs
one-tick durations (Table S2). Some outputs are unstable: they stop
early or loop on a few pitches. Such outputs are part of what the model
produces, so the analysis keeps them. The supplement repeats the
comparisons without them, using the stimulus filter of the listening
test.

\subsubsection{Measures}\label{measures}

Each main measure returns a transposition-invariant score from 0 to 1,
higher for more similar; Table 1 gives its computation. A melody is the
highest note at each onset, the skyline reduction of
\citet{uitdenbogerdMelodicMatchingTechniques1999}. Among about 50
melodic measures compared with expert ratings, n-gram and edit-distance
measures performed best
\citep{mullensiefenOptimizingMeasuresMelodic2004}. Pitch and note-length
histograms are standard features of symbolic music
\citep{tzanetakisPitchHistogramsAudio2002, mckayJSymbolic22Extracting2018, yang2020evaluation}.
The Mallows, or earth mover's, distance best matched human rhythm
judgements among the measures that
\citet{beltranMeasuringMusicalRhythm2013} compared.

Chroma and Duration compute dimensions of CS differently, and Rhythm has
no counterpart in it. Chroma comes closest: over the 3,304 labelled
pairs it ranks pairs much as CS-Chroma does (Spearman 0.80; other main
measures 0.30 to 0.42). CS is reported beside them as a check that the
model learned its own training target. The supplement adds seven
measures of melody, rhythm, register and density, for twelve in all;
three of them are CS or compute a CS component. The four main measures
were fixed after the listening ratings and the first label comparisons,
so the listening analysis is exploratory. Each measure also faces two
human standards in Table 1. The first is the annotation check CS passed:
a measure should score two sections with the same letter above two with
different letters. The second: it should rise with listener ratings.

\begin{table}[!tb]
\centering
\begin{minipage}{\linewidth}
\textbf{Table 1. Melody Agrees Best with Both Human Standards}
\par\medskip
\centering
\renewcommand{\arraystretch}{1.2}
\hyphenpenalty=10000
\begin{tabular}{@{}
  >{\raggedright\arraybackslash}p{(\linewidth - 6\tabcolsep) * \real{0.1236}}
  >{\raggedright\arraybackslash}p{(\linewidth - 6\tabcolsep) * \real{0.5843}}
  >{\raggedleft\arraybackslash}p{(\linewidth - 6\tabcolsep) * \real{0.1461}}
  >{\raggedleft\arraybackslash}p{(\linewidth - 6\tabcolsep) * \real{0.1461}}@{}}
\toprule\noalign{}
\begin{minipage}[b]{\linewidth}\raggedright
Measure
\end{minipage} & \begin{minipage}[b]{\linewidth}\raggedright
Computation
\end{minipage} & \begin{minipage}[b]{\linewidth}\raggedleft
Annotation
\end{minipage} & \begin{minipage}[b]{\linewidth}\raggedleft
Listeners
\end{minipage} \\
\midrule\noalign{}
Melody & Jaccard overlap of the three-interval sequences of the two
melodies & 0.92 & 0.62 \\
Chroma & Cosine similarity of pitch-class histograms, each note counted
once, at the best of 12 transpositions & 0.90 & 0.27 \\
Rhythm & One minus the Mallows distance between inter-onset intervals in
sixteenth notes, capped at one bar, over the cap & 0.75 & 0.23 \\
Duration & Overlap of note-length histograms with eight fixed bins, from
a thirty-second to a double whole note & 0.77 & 0.14 \\
CS (ours) & The label measure & 0.83 & 0.30 \\
\bottomrule
\end{tabular}
\par\medskip
\noindent\emph{Annotation: the probability that a random pair sharing a section
letter outscores a random pair with different letters, over 3,304 pairs
(the annotation check). Listeners: the correlation r with mean ratings
of the 60 clips, on the 10 seconds heard.}
\end{minipage}
\end{table}

Melody clears both standards, with two cautions. The AI letters may
favour melody, since the annotation prompt gives two sections one letter
when they share their primary rhythmic structure and melodic core. On
the 254 pairs both annotators can place, their letters still rank Melody
first: 0.83 for each, against 0.67 and 0.71 for CS. The listener column
comes from the ratings that informed the choice of measures, so it
cannot confirm that choice. The conclusions rest mainly on Melody;
Chroma is close to the training target.

\subsubsection{Analysis}\label{analysis}

For each context and measure, the gap between the scores of two labels
is averaged over the draws. The comparisons are high against low, high
against mid, and mid against low. The mean gap gives the size of an
effect, and the share of contexts with a positive gap its direction.
Contexts from one piece share material, so intervals and tests come from
2,000 resamples of the 49 test pieces, drawn with replacement
\citep{efronBootstrapMethodsAnother1979}. A one-sided p-value is the
share of resamples whose mean gap is zero or below. In every test, the
Holm correction \citep{holmSimpleSequentiallyRejective1979} controls the
chance of any false positive over the four main measures and CS. Table
S3 repeats the main comparisons, corrected over all twelve measures.

\subsection{Results}\label{results}

The results cover what the labels do, what they add over no label, and
two checks.

\subsubsection{Effect of the Labels}\label{effect-of-the-labels}

\begin{figure}
\centering
\includegraphics[width=0.78\linewidth,height=\textheight,keepaspectratio,alt={High-similarity
outputs are closest to the context; mid-similarity outputs are no closer
than low. Each point: share of the 157 contexts, ties left out, where
the first label's output, averaged over draws, is closer. Bars: 95\%
intervals. Panels: high--low, high--mid, mid--low. Filled:
Holm-corrected p \textless{} 0.05 for the mean
gap.}]{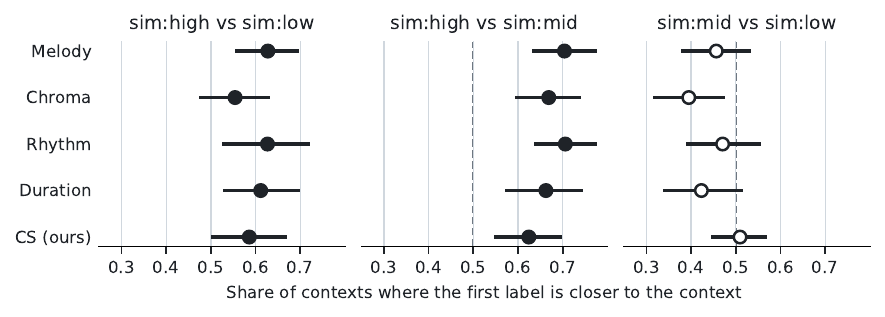}
\caption{High-similarity outputs are closest to the context;
mid-similarity outputs are no closer than low. Each point: share of the
157 contexts, ties left out, where the first label's output, averaged
over draws, is closer. Bars: 95\% intervals. Panels: high--low,
high--mid, mid--low. Filled: Holm-corrected p \textless{} 0.05 for the
mean gap.}
\end{figure}

Every main measure places the high-similarity label above the low and
middle ones (Figure 2). On Melody the mean gap is 0.0080 (95 percent
interval 0.0045--0.0121), and high-similarity outputs come closer in
62.7 percent of contexts. All five remain significant after the Holm
correction over these five, with corrected p-values from 0.003 to 0.012.
All twelve measures of the supplement favour the high label in direction
(Table S3). Table 2 sets the mean gap between the two labels beside the
same gap between real targets.

\begin{table}[!tb]
\centering
\begin{minipage}{\linewidth}
\textbf{Table 2. The Labels Set the Direction, but Cover a Small
Fraction of the Real Difference}
\par\medskip
\centering
\renewcommand{\arraystretch}{1.2}
\hyphenpenalty=10000
\begin{tabular}{@{}lrrrr@{}}
\toprule\noalign{}
Measure & Generated Gap & 95\% Interval & Real Gap & Ratio \\
\midrule\noalign{}
Melody & 0.0080 & 0.0045 to 0.0121 & 0.249 & 3.2\% \\
Chroma & 0.0137 & 0.0048 to 0.0227 & 0.065 & 21.1\% \\
Rhythm & 0.0050 & 0.0019 to 0.0082 & 0.051 & 9.8\% \\
Duration & 0.0154 & 0.0023 to 0.0294 & 0.168 & 9.2\% \\
CS (ours) & 0.0132 & 0.0030 to 0.0231 & 0.355 & 3.7\% \\
\bottomrule
\end{tabular}
\par\medskip
\noindent\emph{Gaps are high minus low in each measure's units, averaged over
draws, then over 157 test contexts. Real gaps compare real high- and
low-similarity targets, cut at the context's length and pooled over
contexts (Table S2).}
\end{minipage}
\end{table}

\begin{figure}
\centering
\includegraphics[width=0.8\linewidth,height=\textheight,keepaspectratio,alt={The
high-similarity output takes up the context's melody; the low-similarity
output hardly does. Six bars of one draw. Red: melody notes, from middle
C up, in a three-interval pattern of the context's melody. Chosen among
the ten largest mean Melody gaps of contexts up to eight bars; typical
gaps are smaller.}]{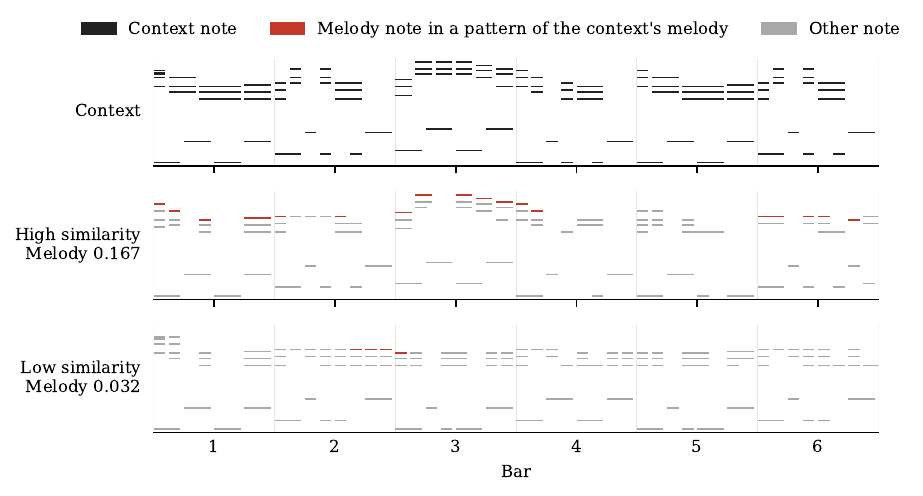}
\caption{The high-similarity output takes up the context's
melody; the low-similarity output hardly does. Six bars of one draw.
Red: melody notes, from middle C up, in a three-interval pattern of the
context's melody. Chosen among the ten largest mean Melody gaps of
contexts up to eight bars; typical gaps are smaller.}
\end{figure}

The generated gap is 3.2\% to 21.1\% of the real gap, and 3.7\% on CS,
the training target. The labels therefore set the direction of
similarity, not its degree. Within a context, single-draw gaps on Melody
have a standard deviation of 0.044, five times the mean gap. One draw at
a time, the high-similarity output is closer on Melody in 54.0\% to
63.0\% of contexts. The data thus show a shift of the mean; six draws
cannot show whether it holds in each context. Figure 3 shows an example
with a clear gap.

A three-level label implies that mid-similarity outputs fall between the
other two. They do not. The high-similarity label exceeds the middle one
on all twelve measures, all after correction, but the middle label
exceeds the low one on none. On Melody, Chroma and Duration, its
interval lies below zero, uncorrected; part of this deficit comes from
unstable outputs (supplement). The pairs are not the cause:
mid-similarity targets are closer to their context than low-similarity
targets on the four main measures (supplement).

Instead, the model compresses the three levels into a narrow band. On
CS, the real targets of the test contexts average 0.918, 0.658 and 0.556
for the high, middle and low labels. The generated outputs average
0.654, 0.638 and 0.641, and those of the no-label control 0.636. All lie
between the real middle and low means. The labels therefore offer
two-way control: the high-similarity label against the rest.

\subsubsection{Label Against No Label}\label{label-against-no-label}

\begin{table}[!tb]
\centering
\begin{minipage}{\linewidth}
\textbf{Table 3. Only the High-Similarity Label Brings Outputs Closer
than No Label Does}
\par\medskip
\centering
\renewcommand{\arraystretch}{1.2}
\hyphenpenalty=10000
\begin{tabular}{@{}
  >{\raggedright\arraybackslash}p{(\linewidth - 10\tabcolsep) * \real{0.1220}}
  >{\raggedleft\arraybackslash}p{(\linewidth - 10\tabcolsep) * \real{0.1707}}
  >{\raggedleft\arraybackslash}p{(\linewidth - 10\tabcolsep) * \real{0.1707}}
  >{\raggedleft\arraybackslash}p{(\linewidth - 10\tabcolsep) * \real{0.1707}}
  >{\raggedleft\arraybackslash}p{(\linewidth - 10\tabcolsep) * \real{0.1707}}
  >{\raggedleft\arraybackslash}p{(\linewidth - 10\tabcolsep) * \real{0.1951}}@{}}
\toprule\noalign{}
\begin{minipage}[b]{\linewidth}\raggedright
Measure
\end{minipage} & \begin{minipage}[b]{\linewidth}\raggedleft
High \textgreater{} No Label
\end{minipage} & \begin{minipage}[b]{\linewidth}\raggedleft
No Label \textgreater{} Low
\end{minipage} & \begin{minipage}[b]{\linewidth}\raggedleft
No Label \textgreater{} Mid
\end{minipage} & \begin{minipage}[b]{\linewidth}\raggedleft
Wrong Label: High \textgreater{} Low
\end{minipage} & \begin{minipage}[b]{\linewidth}\raggedleft
No Label \textgreater{} Unconditioned
\end{minipage} \\
\midrule\noalign{}
Melody & \textbf{66\% (0.003)} & 49\% (1.000) & 54\% (0.993) &
\textbf{36\% (0.020)} & 57\% (0.453) \\
Chroma & \textbf{58\% (0.024)} & 54\% (0.922) & \textbf{59\% (0.016)} &
\textbf{45\% (0.044)} & \textbf{68\% (0.003)} \\
Rhythm & \textbf{69\% (0.003)} & 49\% (1.000) & 53\% (0.993) & 52\%
(1.000) & 57\% (0.196) \\
Duration & 53\% (0.067) & 54\% (0.923) & \textbf{65\% (0.005)} & 46\%
(0.231) & \textbf{62\% (0.010)} \\
CS (ours) & \textbf{60\% (0.004)} & 44\% (1.000) & 49\% (0.993) & 50\%
(1.000) & 50\% (0.214) \\
\bottomrule
\end{tabular}
\par\medskip
\noindent\emph{Share of 157 contexts where a column's first output is closer to
the context. In brackets: the mean gap's p, Holm-corrected per column,
one-sided except the mid and wrong labels. Bold: p \textless{} 0.05.
Ties are left out.}
\end{minipage}
\end{table}

The no-label control shares the context vector and lacks only the label,
so it shows what the label adds (Table 3). The high-similarity label
brings outputs closer than no label on every measure but Duration. On
Melody it does so in 66.2 percent of contexts, by 0.0106 (95 percent
interval 0.0063--0.0156). The low-similarity label does not move outputs
measurably. On Melody, low-labelled outputs lie 0.0026 closer than
no-label ones (-0.0010 to +0.0064), an interval that excludes the high
label's shift. The middle label moves them further away on Chroma and
Duration, partly through unstable outputs (Table S4). Meaning matters
too: the wrong-label control's high-similarity label always marked a mid
or low pair, and its order reverses on Melody and Chroma.

Selection is the obvious alternative: generate twelve no-label outputs
and keep the closest to the context by CS. On Melody, Rhythm and
Duration, it cannot be told apart from one high-similarity output; on
Melody the gap is 0.0008 (-0.0066 to +0.0082). Selection comes closer on
Chroma.

Against the unconditioned control, the context vector raises Chroma and
Duration but not Melody or Rhythm (Table 3). It also halves unstable
outputs, from 30.6\% to 14.1\%, which may explain part of these gains;
its Chroma gain survives the stimulus filter (Table S4). The context
vector steadies outputs and adds pitch content, and the label adds
melody.

\subsubsection{Context Specificity and
Restatement}\label{context-specificity-and-restatement}

If the label only made outputs more typical of the corpus, the
high-minus-low gap would be as large against contexts from other pieces.
We therefore scored every output against three foreign contexts drawn at
random from other pieces. Pieces from one game can share a theme, which
makes the check conservative. The gap is not as large. On Melody, it is
0.0080 against the own context and -0.0001 against foreign contexts, a
difference of 0.0081 (95 percent interval 0.0041--0.0126). Chroma,
Rhythm and CS agree, and all four differences survive the Holm
correction (Table S5). On Duration the difference, 0.0103, has an
interval that includes zero. This shows the label brings melody, pitch
content and rhythm closer to the specific context; on note lengths it is
inconclusive.

Composers vary a theme by restating parts of it; a model asked for high
similarity may do the same, or merely duplicate its context. Over all
draws, we counted outputs whose top voice restates a run of eight or
more notes of the context's melody, in transposition-invariant
intervals. Restatement is most frequent under the high-similarity label,
in 12.4\% of outputs (95\% interval 9.0--16.4\%). Mid-, low-similarity
and no-label outputs restate in 6.0\%, 7.1\% and 7.0\%. Real targets of
equal CS, cut to the context's length, restate in 17.4\% (9.7--27.5\%),
so the outputs show no excess. Nor do they duplicate the context:
restatement fills most of 1.0\% of high outputs, against 32.4\% of real
high targets.

On melody, restatement is how the label acts. Without restating outputs,
the Melody gap falls to 0.0027, with an interval reaching zero; Chroma
and Rhythm keep theirs (supplement). Outputs do not reproduce other
pieces. No test context scores 0.5 or more on Melody against any
training section. Without the 39 contexts that share an eight-note run
with one, the Melody gap is 0.0078 (0.0038--0.0122).

\section{Listening Test}\label{listening-test}

The test asks two questions. Does a high-similarity output sound more
similar to its context than the low-similarity output? And which
measures match what listeners hear?

\subsection{Design}\label{design-1}

The design covers the stimuli, the procedure with its listeners, and the
analysis.

\subsubsection{Stimuli}\label{stimuli}

The stimuli are first-draw outputs of SimLlama, one high- and one
low-similarity output per context. Each trial presents one draw per
label, and an unstable clip would turn a similarity rating into a
quality rating. A stimulus filter therefore admitted a context only if
neither output was unstable, treating both labels alike (thresholds in
the supplement). Thirty contexts were then drawn at random, with five
replacements (supplement). All clips, the context included, were
rendered as solo piano and cut after 10 seconds, so one session could
hold 30 contexts.

\subsubsection{Procedure and Listeners}\label{procedure-and-listeners}

Each trial presented three replayable clips: the context and two
sections, A and B. Listeners rated how similar A and B each were to the
context, on two sliders from 0, completely dissimilar, to 100,
completely similar. Listeners, blind to the labels, were unpaid
volunteers invited by email who took the test online.

The session held the 30 test trials and two attention checks in one
shuffled order and A/B assignment, the same for every listener. The
high-similarity output was section A in half of the trials. Each check
pairs a real section of the same piece with a section from an unrelated
piece. A listener who rated the unrelated section higher in either check
was excluded, a rule fixed in advance. The study was approved by the
Queen Mary University of London research ethics committee (reference
DSEECS26.094) and recorded no names or contact details. Twenty-six
listeners completed the session, and six failed an attention check,
leaving 20.

\subsubsection{Analysis}\label{analysis-1}

Both questions use linear mixed models with crossed random intercepts
for listener and context \citep{baayenMixedEffectsModeling2008}. The
label effect, the gap between a trial's two ratings, is modelled with an
intercept and a slot term. The slot term is +1 when the high-similarity
output came first and -1 otherwise. Measure agreement predicts each
rating from the measure's standardised score for that clip and a slot
term, with a further random intercept for the clip. Each measure is
computed on the 10 seconds heard. With twelve measures compared on one
test, each p-value comes with its Holm correction.

\subsection{Results}\label{results-1}

The results answer the two questions in turn.

\subsubsection{Label Effect}\label{label-effect}

The test did not detect the label effect. Listeners rated
high-similarity outputs 2.9 points higher (95 percent interval -2.0 to
+7.7, p = 0.24). This model generalises over trials as well as
listeners; a test over listeners alone, which holds the 30 trials fixed,
is significant (supplement). With all 26 listeners it is 2.2 points
(-1.8 to +6.2).

The stimuli help explain this. Each trial played a single draw per
label, and single draws scatter widely. On the 10 seconds heard, the 30
pairs show no objective gap: the high-similarity output shares more
melody in only 12. Over all passing draws, the same 10 seconds favour it
on Melody in 67.1 percent of the 157 contexts; the single draws heard do
not. With 20 listeners and 30 trials, the test had 80 percent power only
for an effect of about 7 points.

Listeners also favoured section A, played straight after the context, by
7.9 points whatever it held (95 percent interval 3.1--12.6, p = 0.001).
The slot terms adjust for it.

\subsubsection{Measure Agreement}\label{measure-agreement}

\begin{table}[!tb]
\centering
\begin{minipage}{\linewidth}
\textbf{Table 4. Of the Main Measures, Melody Relates Most to Listener
Ratings}
\par\medskip
\centering
\renewcommand{\arraystretch}{1.2}
\hyphenpenalty=10000
\begin{tabular}{@{}lrrrr@{}}
\toprule\noalign{}
Measure & Rise per SD & 95\% Interval & p & Holm p \\
\midrule\noalign{}
Melody & +6.7 & 4.6 to 8.9 & \textless{} 0.001 & \textless{} 0.001 \\
Chroma & +2.9 & 0.2 to 5.5 & 0.035 & 0.21 \\
Rhythm & +2.7 & -0.04 to 5.4 & 0.054 & 0.22 \\
Duration & +1.5 & -1.2 to 4.3 & 0.28 & 0.44 \\
CS (ours) & +3.4 & 0.9 to 6.0 & 0.009 & 0.060 \\
\bottomrule
\end{tabular}
\par\medskip
\noindent\emph{20 listeners, 60 clips of 10 seconds. Mixed model with a slot term
and random intercepts for listener, context and clip; Holm correction
over the twelve measures of Table S6.}
\end{minipage}
\end{table}

Only Melody survives the Holm correction (Table 4). CS narrowly misses
after it, and Chroma passes only before it. Over all 12, two further
melodic measures survive, as do Register, which compares mean pitch, and
Density, which compares note rates (Table S6). The two melodic measures
read the same melody, and a joint model separates the other two. With
Melody, Density, Register and CS entered together, Melody keeps a rise
of 5.9 points (3.2--8.7); the other three lose their effect. In this
model, Melody exceeds CS by 5.4 points (1.1--9.7); without the four
restating clips below, its lead over CS is not significant (supplement).
Adding Melody to the other three improves the model clearly (likelihood
ratio 25.0, p \textless{} 0.001).

Part of this effect comes from a few clips. Four of 60 clips restate
eight or more notes of their context, two under each label. Without
them, Melody's rise falls to 3.6 points (1.4--5.8), still significant,
and two further melodic measures and Density rank above it, up to 4.1.
Melody is also zero for 28 clips, which share no three-interval pattern
with their context. Figure 4 shows the pattern trial by trial: the
Melody gap ranks the listener gap with Spearman \(\rho = 0.55\) (p =
0.002).

\begin{figure}
\centering
\includegraphics[width=0.46\linewidth,height=\textheight,keepaspectratio,alt={Listeners
hear the output that shares more melody with the context as more
similar. Each point is one trial: the Melody gap between its two outputs
on the 10 seconds heard, against the mean listener gap without the slot
bias. Bars are 95\% intervals across 20
listeners.}]{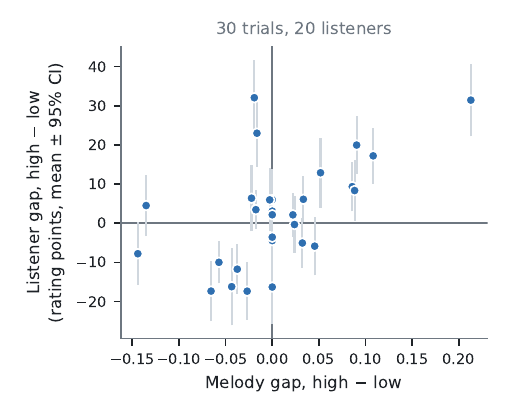}
\caption{Listeners hear the output that shares more melody with
the context as more similar. Each point is one trial: the Melody gap
between its two outputs on the 10 seconds heard, against the mean
listener gap without the slot bias. Bars are 95\% intervals across 20
listeners.}
\end{figure}

\section{Discussion}\label{discussion}

The findings show what the labels can do and how the label measure can
improve.

\subsection{Main Findings}\label{main-findings}

The labels set the direction of similarity, specific to each context;
they do not yet set its degree, and act on average. Against no label,
only the high-similarity label makes outputs more similar. On melody it
does so mainly by restating the context's melody more often. For a
composer, one labelled output could not be told apart from the closest
of twelve no-label outputs on Melody, Rhythm and Duration. On Chroma,
selection came closer.

\subsection{Design of CS}\label{design-of-cs}

Three choices made before any listening data explain part of what the
model learned. First, CS leaves out melody. At design time we judged
that much game music has no clear melody. We also judged that CS-Chroma
and CS-Register capture a melody's pitch content and range, though not
its order. Table 1 does not support the second judgement. A skyline
melody also misses inner and bass lines, and in piano reductions
accompaniment above the tune breaks it. The Melody measure inherits
these errors. A melodic component on separated voices is the next step.

Second, the weights were set a priori, since no published weights exist
for these components. CS-Chroma alone separates the annotated pairs
better than CS (0.873 against 0.834; 95 percent interval of the
difference 0.005 to 0.078). Weights fitted to the annotated pairs and
tested on listeners may do better.

Third, tertile thresholds split one dense peak into the low and middle
labels, while the high label spans a broad range (Figure 1). It is also
a loose target: only 38.2 percent of high-labelled pairs share a section
letter. This may explain why only the high label separates. Other
thresholds, two labels or a continuous value would test this.

\subsection{Further Limits}\label{further-limits}

All pieces are piano transcriptions from one game series. The model
continues a section freely, without control of loop points, harmonic
boundaries or key, which the measures also ignore. Each model is one
training run. A study with composers would assess the outputs' musical
quality and usefulness.

\section{Conclusion}\label{conclusion}

On average, high-similarity outputs lie closer to their context than
low-similarity ones, in direction on all twelve measures. Against no
label, only the high-similarity label makes outputs more similar; on
melody, mainly through more frequent restatement of the context's
melody. The shift is small: on CS, all outputs stay between the real
middle and low means. The context vector halves unstable outputs.

Listeners did not detect the label effect in single 10-second draws.
Their ratings followed melodic overlap most closely of the measures
tested, which the label measure does not see. The ratings thus point to
a concrete next step: a label measure that includes melody.

\section{Code and Data Availability}\label{code-and-data-availability-1}

\begin{itemize}
\tightlist
\item
  Code: \url{https://github.com/ShxLuo-Saxon/simllama},
  doi:10.5281/zenodo.23915905
\item
  Structure annotations and data:
  \url{https://github.com/ShxLuo-Saxon/supermario-structure-dataset},
  doi:10.5281/zenodo.23915907
\item
  Model checkpoint:
  \url{https://huggingface.co/shangxuan-saxon/SimLlama}
\item
  Audio examples: \url{https://shxluo-saxon.github.io/simllama/}
\end{itemize}

\section{Acknowledgments}
This work was supported by the Queen Mary University of London and China
Scholarship Council (CSC) Joint Scholarship (grant 202408060136). We
thank the listening-test participants and the annotator who provided the
second structure annotation.
\bibliographystyle{cmj}
\bibliography{refs}

\begin{thebibliography}{43}
\newcommand{\enquote}[1]{``#1''}
\providecommand{\natexlab}[1]{#1}
\providecommand{\url}[1]{\texttt{#1}}
\providecommand{\urlprefix}{Available online at }

\bibitem[{Amaral et~al.(2022)Amaral, Baffa, Briot, Feijó, and
  Furtado}]{amaral2022adaptive}
Amaral, G., et~al. 2022.
\newblock \enquote{An adaptive music generation architecture for games based on
  the deep learning Transformer model.}
\newblock In \emph{21st Brazilian Symposium on Computer Games and Digital
  Entertainment (SBGames)}, pp. 1--6.

\bibitem[{Baayen, Davidson, and Bates(2008)}]{baayenMixedEffectsModeling2008}
Baayen, R.~H., D.~J. Davidson, and D.~M. Bates. 2008.
\newblock \enquote{Mixed-effects modeling with crossed random effects for
  subjects and items.}
\newblock \emph{Journal of Memory and Language} 59(4):390--412.

\bibitem[{Beltran et~al.(2013)Beltran, Liu, Mohanchandra, and
  Toussaint}]{beltranMeasuringMusicalRhythm2013}
Beltran, J.~F., et~al. 2013.
\newblock \enquote{Measuring Musical Rhythm Similarity - Statistical Features
  versus Transformation Methods.}
\newblock In \emph{Proceedings of the 2nd International Conference on Pattern
  Recognition Applications and Methods}, pp. 595--598.

\bibitem[{Bhandari et~al.(2025)Bhandari, Chang, Lu, Enus, Bradshaw, Herremans,
  and Colton}]{bhandariImprovNetGeneratingControllable2025}
Bhandari, K., et~al. 2025.
\newblock \enquote{ImprovNet -- Generating Controllable Musical Improvisations
  with Iterative Corruption Refinement.}
\newblock Available online at \url{arxiv.org/abs/2502.04522}. Accessed November
  2026.

\bibitem[{Cohen(1960)}]{cohenCoefficientAgreementNominal1960}
Cohen, J. 1960.
\newblock \enquote{A Coefficient of Agreement for Nominal Scales.}
\newblock \emph{Educational and Psychological Measurement} 20(1):37--46.

\bibitem[{Dai et~al.(2021)Dai, Jin, Gomes, and
  Dannenberg}]{daiControllableDeepMelody2021}
Dai, S., et~al. 2021.
\newblock \enquote{Controllable deep melody generation via hierarchical music
  structure representation.}
\newblock In \emph{Proceedings of the 22nd International Society for Music
  Information Retrieval Conference}, pp. 143--150.

\bibitem[{Deliège(2001)}]{deliegeSimilarityPerceptionCategorization2001}
Deliège, I. 2001.
\newblock \enquote{Similarity Perception {$\leftrightarrow$} Categorization
  {$\leftrightarrow$} Cue Abstraction.}
\newblock \emph{Music Perception} 18(3):233--243.

\bibitem[{Eerola et~al.(2001)Eerola, Järvinen, Louhivuori, and
  Toiviainen}]{eerolaStatisticalFeaturesPerceived2001}
Eerola, T., et~al. 2001.
\newblock \enquote{Statistical Features and Perceived Similarity of Folk
  Melodies.}
\newblock \emph{Music Perception} 18(3):275--296.

\bibitem[{Efron(1979)}]{efronBootstrapMethodsAnother1979}
Efron, B. 1979.
\newblock \enquote{Bootstrap Methods: Another Look at the Jackknife.}
\newblock \emph{The Annals of Statistics} 7(1):1--26.

\bibitem[{Ferreira, Lelis, and Whitehead(2020)}]{ferreira2020computergenerated}
Ferreira, L., L.~Lelis, and J.~Whitehead. 2020.
\newblock \enquote{Computer-Generated Music for Tabletop Role-Playing Games.}
\newblock \emph{Proceedings of the AAAI Conference on Artificial Intelligence
  and Interactive Digital Entertainment} 16(1):59--65.

\bibitem[{Gao, Reuben, and Collins(2024)}]{gaoVariationTransformerNew2024}
Gao, C., F.~Reuben, and T.~Collins. 2024.
\newblock \enquote{Variation Transformer: New Datasets, Models, and Comparative
  Evaluation for Symbolic Music Variation Generation.}
\newblock In \emph{Proceedings of the 25th International Society for Music
  Information Retrieval Conference}, pp. 154--163.

\bibitem[{{Gemini Team} et~al.(2023){Gemini Team}, Anil, Borgeaud, Alayrac
  et~al.}]{teamGeminiFamilyHighly2025}
{Gemini Team}, et~al. 2023.
\newblock \enquote{Gemini: A Family of Highly Capable Multimodal Models.}
\newblock Available online at \url{arxiv.org/abs/2312.11805}. Accessed November
  2026.

\bibitem[{Guo and Dixon(2025)}]{guo2025moonbeam}
Guo, Z., and S.~Dixon. 2025.
\newblock \enquote{Moonbeam: A MIDI Foundation Model Using Both Absolute and
  Relative Music Attributes.}
\newblock Available online at \url{arxiv.org/abs/2505.15559}. Accessed November
  2026.

\bibitem[{Holm(1979)}]{holmSimpleSequentiallyRejective1979}
Holm, S. 1979.
\newblock \enquote{A Simple Sequentially Rejective Multiple Test Procedure.}
\newblock \emph{Scandinavian Journal of Statistics} 6(2):65--70.

\bibitem[{Hu et~al.(2022)Hu, Shen, Wallis, Allen-Zhu, Li, Wang, Wang, and
  Chen}]{huLoRALowRankAdaptation2021a}
Hu, E.~J., et~al. 2022.
\newblock \enquote{LoRA: Low-Rank Adaptation of Large Language Models.}
\newblock In \emph{International Conference on Learning Representations}.
\newblock Available online at \url{openreview.net/forum?id=nZeVKeeFYf9}.
  Accessed November 2026.

\bibitem[{Ji, Yang, and Luo(2023)}]{ji2024survey}
Ji, S., X.~Yang, and J.~Luo. 2023.
\newblock \enquote{A Survey on Deep Learning for Symbolic Music Generation:
  Representations, Algorithms, Evaluations, and Challenges.}
\newblock \emph{ACM Computing Surveys} 56(1):1--39.

\bibitem[{Lamont and Dibben(2001)}]{lamontMotivicStructurePerception2001}
Lamont, A., and N.~Dibben. 2001.
\newblock \enquote{Motivic Structure and the Perception of Similarity.}
\newblock \emph{Music Perception} 18(3):245--274.

\bibitem[{Landis and Koch(1977)}]{landisMeasurementObserverAgreement1977}
Landis, J.~R., and G.~G. Koch. 1977.
\newblock \enquote{The Measurement of Observer Agreement for Categorical Data.}
\newblock \emph{Biometrics} 33(1):159--174.

\bibitem[{Lerch et~al.(2026)Lerch, Arthur, Bryan-Kinns, Ford, Sun, and
  Vinay}]{lerch2026survey}
Lerch, A., et~al. 2026.
\newblock \enquote{Survey on the Evaluation of Generative Models in Music.}
\newblock \emph{ACM Computing Surveys} 58(4):1--36.

\bibitem[{Lopez~Duarte(2024)}]{lopezduarte2024progressiveadaptive}
Lopez~Duarte, A. 2024.
\newblock \enquote{A Progressive-Adaptive Music Generator (PAMG): An Approach
  to Interactive Procedural Music for Videogames.}
\newblock In \emph{Proceedings of the 12th ACM SIGPLAN International Workshop
  on Functional Art, Music, Modelling, and Design}, pp. 65--72.

\bibitem[{Luo and Reiss(2025)}]{shangxuanluo2025procedural}
Luo, S., and J.~Reiss. 2025.
\newblock \enquote{Procedural Music Generation Systems in Games.}
\newblock In \emph{AES International Conference on Machine Learning and
  Artificial Intelligence for Audio}.

\bibitem[{McIntosh, Woscholski, and
  Barthet(2023)}]{mcintoshAffectiveConditionalModifiers2023}
McIntosh, T.~H., O.~Woscholski, and M.~Barthet. 2023.
\newblock \enquote{Affective Conditional Modifiers in Adaptive Video Game
  Music.}
\newblock In \emph{Proceedings of the 18th International Audio Mostly
  Conference}, pp. 17--23.

\bibitem[{McKay, Cumming, and Fujinaga(2018)}]{mckayJSymbolic22Extracting2018}
McKay, C., J.~E. Cumming, and I.~Fujinaga. 2018.
\newblock \enquote{jSymbolic 2.2: Extracting Features from Symbolic Music for
  Use in Musicological and MIR Research.}
\newblock In \emph{Proceedings of the 19th International Society for Music
  Information Retrieval Conference}, pp. 348--354.

\bibitem[{Müllensiefen and
  Frieler(2004)}]{mullensiefenOptimizingMeasuresMelodic2004}
Müllensiefen, D., and K.~Frieler. 2004.
\newblock \enquote{Optimizing Measures of Melodic Similarity for the
  Exploration of a Large Folk Song Database.}
\newblock In \emph{Proceedings of the 5th International Conference on Music
  Information Retrieval (ISMIR)}.
\newblock Available online at
  \url{archives.ismir.net/ismir2004/paper/000178.pdf}. Accessed November 2026.

\bibitem[{Phillips(2014)}]{winifred2017composers}
Phillips, W. 2014.
\newblock \emph{A Composer's Guide to Game Music}.
\newblock MIT Press.

\bibitem[{Plut and Pasquier(2020)}]{plut2020generative}
Plut, C., and P.~Pasquier. 2020.
\newblock \enquote{Generative music in video games: State of the art,
  challenges, and prospects.}
\newblock \emph{Entertainment Computing} 33:100337.

\bibitem[{Roberts et~al.(2018)Roberts, Engel, Raffel, Hawthorne, and
  Eck}]{roberts2018hierarchical}
Roberts, A., et~al. 2018.
\newblock \enquote{A Hierarchical Latent Vector Model for Learning Long-Term
  Structure in Music.}
\newblock In \emph{Proceedings of the 35th International Conference on Machine
  Learning}, \emph{Proceedings of Machine Learning Research}, vol.~80, pp.
  4364--4373.

\bibitem[{Rütte et~al.(2023)Rütte, Biggio, Kilcher, and
  Hofmann}]{rutteFIGAROGeneratingSymbolic2024}
Rütte, D.~v., et~al. 2023.
\newblock \enquote{FIGARO: Controllable Music Generation using Expert and
  Learned Features.}
\newblock In \emph{International Conference on Learning Representations}.
\newblock Available online at \url{arxiv.org/abs/2201.10936}. Accessed November
  2026.

\bibitem[{Shih et~al.(2023)Shih, Wu, Zalkow, Müller, and
  Yang}]{shihThemeTransformerSymbolic2023}
Shih, Y.-J., et~al. 2023.
\newblock \enquote{Theme Transformer: Symbolic Music Generation With
  Theme-Conditioned Transformer.}
\newblock \emph{IEEE Transactions on Multimedia} 25:3495--3508.

\bibitem[{Summers(2016)}]{summers2016understanding}
Summers, T. 2016.
\newblock \emph{Understanding video game music}.
\newblock Cambridge University Press.

\bibitem[{Tan and Herremans(2020)}]{tan2020music}
Tan, H.~H., and D.~Herremans. 2020.
\newblock \enquote{Music FaderNets: Controllable Music Generation Based On
  High-Level Features via Low-Level Feature Modelling.}
\newblock In \emph{Proceedings of the 21st International Society for Music
  Information Retrieval Conference}, pp. 109--116.

\bibitem[{Tanabe et~al.(2023)Tanabe, Khan, Nguyen, Nimpattanavong, and
  Thawonmas}]{tanabe2023adaptive}
Tanabe, J., et~al. 2023.
\newblock \enquote{Adaptive Background Music According to the Player’s
  Arousal for DareFightingICE.}
\newblock In \emph{Proceedings of the 13th International Conference on Advances
  in Information Technology}, pp. 1--6.

\bibitem[{Thickstun et~al.(2024)Thickstun, Hall, Donahue, and
  Liang}]{thickstunAnticipatoryMusicTransformer2023}
Thickstun, J., et~al. 2024.
\newblock \enquote{Anticipatory Music Transformer.}
\newblock \emph{Transactions on Machine Learning Research} \unskip. Available
  online at \url{openreview.net/forum?id=EBNJ33Fcrl}. Accessed November 2026.

\bibitem[{Touvron et~al.(2023)Touvron, Lavril, Izacard, Martinet
  et~al.}]{touvronLLaMAOpenEfficient2023}
Touvron, H., et~al. 2023.
\newblock \enquote{LLaMA: Open and Efficient Foundation Language Models.}
\newblock Available online at \url{arxiv.org/abs/2302.13971}. Accessed November
  2026.

\bibitem[{Tzanetakis, Ermolinskyi, and
  Cook(2002)}]{tzanetakisPitchHistogramsAudio2002}
Tzanetakis, G., A.~Ermolinskyi, and P.~R. Cook. 2002.
\newblock \enquote{Pitch Histograms in Audio and Symbolic Music Information
  Retrieval.}
\newblock In \emph{Proceedings of the 3rd International Conference on Music
  Information Retrieval}.
\newblock Available online at \url{doi.org/10.5281/zenodo.1416145}. Accessed
  November 2026.

\bibitem[{Uitdenbogerd and
  Zobel(1999)}]{uitdenbogerdMelodicMatchingTechniques1999}
Uitdenbogerd, A., and J.~Zobel. 1999.
\newblock \enquote{Melodic matching techniques for large music databases.}
\newblock In \emph{Proceedings of the seventh ACM international conference on
  Multimedia (Part 1)}, pp. 57--66.

\bibitem[{Volk, de~Haas, and van Kranenburg(2012)}]{volk2012modelling}
Volk, A., W.~B. de~Haas, and P.~van Kranenburg. 2012.
\newblock \enquote{Towards Modelling Variation in Music as Foundation for
  Similarity.}
\newblock In E.~Cambouropoulos, C.~Tsougras, P.~Mavromatis, and K.~Pastiadis,
  eds. \emph{Proceedings of the 12th International Conference on Music
  Perception and Cognition and the 8th Triennial Conference of the European
  Society for the Cognitive Sciences of Music}. School of Music Studies,
  Aristotle University of Thessaloniki, pp. 1085--1094.

\bibitem[{Wang et~al.(2025)Wang, Wu, Hu, Du et~al.}]{wang2025notagen}
Wang, Y., et~al. 2025.
\newblock \enquote{NotaGen: Advancing Musicality in Symbolic Music Generation
  with Large Language Model Training Paradigms.}
\newblock In \emph{Proceedings of the Thirty-Fourth International Joint
  Conference on Artificial Intelligence}, pp. 10207--10215.

\bibitem[{Wang, Min, and Xia(2024)}]{wangWholeSongHierarchical2024}
Wang, Z., L.~Min, and G.~Xia. 2024.
\newblock \enquote{Whole-Song Hierarchical Generation of Symbolic Music Using
  Cascaded Diffusion Models.}
\newblock In \emph{International Conference on Learning Representations}.
\newblock Available online at \url{arxiv.org/abs/2405.09901}. Accessed November
  2026.

\bibitem[{Worrall(2025)}]{worrallExpressivePerformanceRendering2025}
Worrall, K. 2025.
\newblock \enquote{Expressive Performance Rendering for Reduced Music
  Repetition in Video Games.}
\newblock PhD dissertation, University of York, York, UK.

\bibitem[{Worrall and Collins(2024)}]{worrall2023considerations}
Worrall, K., and T.~Collins. 2024.
\newblock \enquote{Considerations and Concerns of Professional Game Composers
  Regarding Artificially Intelligent Music Technology.}
\newblock \emph{IEEE Transactions on Games} 16(3):586--597.

\bibitem[{Wu and Yang(2023)}]{wuMuseMorphoseFullSong2021}
Wu, S.-L., and Y.-H. Yang. 2023.
\newblock \enquote{MuseMorphose: Full-Song and Fine-Grained Piano Music Style
  Transfer with One Transformer VAE.}
\newblock \emph{IEEE/ACM Transactions on Audio, Speech, and Language
  Processing} 31:1953--1967.

\bibitem[{Yang and Lerch(2020)}]{yang2020evaluation}
Yang, L.-C., and A.~Lerch. 2020.
\newblock \enquote{On the evaluation of generative models in music.}
\newblock \emph{Neural Computing and Applications} 32(9):4773--4784.

\end{thebibliography}

\clearpage
\setcounter{figure}{0}
\renewcommand{\thefigure}{S\arabic{figure}}
\FloatBarrier
\section{Supplementary Material}\label{supp-supplementary-material}

The supplement follows the order of the main text. It covers the
validation procedure, the formulas of every measure, the model input and
the details needed for reproduction. It then gives the label comparisons
with further checks, the stimulus filter and the control models, context
specificity, and the listening analyses.

\FloatBarrier
\subsection{Validation Procedure}\label{supp-validation-procedure}

Figure S1 shows one piece annotated both ways. The AI gave a section the
letter of an earlier one if it ``shares the same primary rhythmic
structure and melodic core''. The prompt allowed slight variations in
harmony or instrumentation. A new letter marked a section that
``introduces a fundamentally new musical idea'', a contrasting rhythmic
motif, or a significant shift in texture. The comparison with the human
hierarchies counts bars from the first bar of each annotation. For each
AI section and each human level, it first finds whether the section
starts and ends on boundaries of that level. If so, the section equals
one human segment there, or a run of consecutive segments, and covers
their letters. An AI section is consistent with an annotator if this
holds at some level.

\begin{figure}
\centering
\includegraphics[width=0.85\linewidth,height=\textheight,keepaspectratio,alt={The
AI marks one level of sections; the annotators mark nested levels. One
piece, ``Title Screen'', annotated by the AI and both annotators. Each
row is one level, and letters compare segments within a row only. The
AI's sections match the first annotator's second, third, second and
second level. The annotators disagree on whether the last section
repeats the second.}]{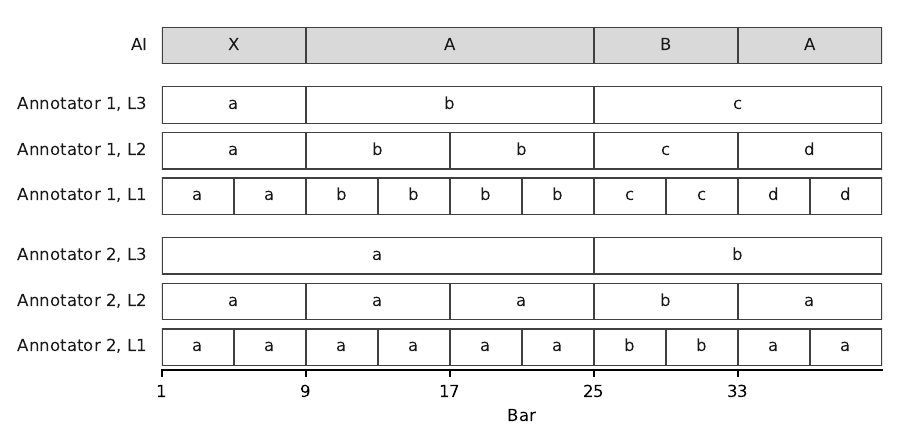}
\caption{The AI marks one level of sections; the annotators mark
nested levels. One piece, ``Title Screen'', annotated by the AI and
both annotators. Each row is one level, and letters compare segments
within a row only. The AI's sections match the first annotator's second,
third, second and second level. The annotators disagree on whether the
last section repeats the second.}
\end{figure}

For each pair of AI sections in a piece, the comparison takes the lowest
level at which both equal one segment. Failing that, it takes the lowest
at which both hold. The annotator judges the pair the same material if
the two letter sets are equal there. The AI judges a pair by its own
letters. The two schemes label material outside the main body
differently. Pairs are therefore left out if a section is the AI's X.
The prompt gives X to an introduction, ending, bridge or stinger; a
bridge links sections without restating their material. They are also
left out if a section covers an annotator's x for silence.

The validation uses all 47 pieces. The annotation check of CS uses only
the labelled pairs. It covers the 30 of these pieces that are among the
334 labelled pieces and that both annotators can place. Agreement and
Cohen's kappa are computed between each two sources over all pairs that
both annotators judge, with intervals from 2,000 resamples of whole
pieces. The annotation check pools same-letter and different-letter
pairs over all pieces and counts a tie as half a win, as the
Mann--Whitney statistic does.

\FloatBarrier
\subsection{Measures}\label{supp-measures}

Every measure compares an output \(b\) with its context \(a\) and
returns a score between 0 and 1, where higher means more similar. All
but CS-Register, Register and CS, which contains CS-Register, are
invariant to transposition. A melody is the highest note at each onset.
Melody compares the sets of distinct three-interval sequences and scores
zero when either set is empty. The rhythm measures merge notes struck
together into one onset and cap inter-onset intervals at 16 sixteenth
notes. An overlap of two histograms is one minus half the summed
absolute difference of their normalised bins. Table S1 gives each
formula.

\begingroup\singlespacing\small
\renewcommand{\arraystretch}{1.2}\hyphenpenalty=10000
\begin{longtable}[]{@{}
  >{\raggedright\arraybackslash}p{(\linewidth - 2\tabcolsep) * \real{0.5000}}
  >{\raggedright\arraybackslash}p{(\linewidth - 2\tabcolsep) * \real{0.5000}}@{}}
\multicolumn{2}{@{}p{\linewidth}@{}}{\normalsize \textbf{Table S1. Formulas of the Label Measure's Components and of the
Seven Additional Measures}}\\[6pt]
\toprule\noalign{}
\begin{minipage}[b]{\linewidth}\raggedright
Name
\end{minipage} & \begin{minipage}[b]{\linewidth}\raggedright
Computation
\end{minipage} \\
\midrule\noalign{}
\endfirsthead
\toprule\noalign{}
\begin{minipage}[b]{\linewidth}\raggedright
Name
\end{minipage} & \begin{minipage}[b]{\linewidth}\raggedright
Computation
\end{minipage} \\
\midrule\noalign{}
\endhead
\bottomrule\noalign{}
\endlastfoot
CS-Chroma & Highest cosine similarity of the duration-weighted
pitch-class histograms over the 12 transpositions of \(b\) \\
CS-Duration & Cosine similarity of note-length histograms with 16
logarithmic bins between each section's own shortest and longest note,
so the two sections' bins differ \\
CS-Register & \(\max(0, 1 - \lvert \bar{p}_a - \bar{p}_b \rvert / 48)\),
with \(\bar{p}\) the mean pitch in semitones \\
CS-Density & \(1 - \lvert r_a - r_b \rvert / \max(r_a, r_b)\), with
\(r\) the notes per unit time \\
CS & \((4 c_1 + 3 c_2 + c_3 + c_4)/9\) for the four components in this
order, the weights renormalised when a component cannot be computed \\
Melody-Contour & One minus the edit distance of the up, down and same
strings of the two melodies, at most 64 steps, over the longer string's
length \\
Melody-Intervals & Overlap of melodic-interval histograms from -12 to
+12 semitones \\
Melody-Alignment & \(e^{-c/n}\), with \(c\) the cost of a
dynamic-time-warping alignment of intervals folded into one octave, each
step costing the absolute interval difference, and \(n\) the summed
length of the two sequences \\
Rhythm-Overlap & Overlap of inter-onset-interval histograms in sixteenth
notes \\
Rhythm-Grid & Cosine similarity of onset counts at the sixteen
sixteenth-note positions of a four-beat cycle \\
Register & \(\max(0, 1 - \lvert \bar{p}_a - \bar{p}_b \rvert / 88)\); it
ranks pairs as CS-Register does, on the scale of the piano keyboard \\
Density & The same computation as CS-Density, reported so that every
dimension appears among the evaluation measures \\
\end{longtable}
\par\endgroup

\FloatBarrier
\subsection{Model Input and
Reproducibility}\label{supp-model-input-and-reproducibility}

Figure S2 draws the input of SimLlama, and Table S2 lists the details
needed to reproduce the results.

\begin{figure}
\centering
\includegraphics[width=0.85\linewidth,height=\textheight,keepaspectratio,alt={A
label and a context vector condition a frozen pretrained model. Context,
label and target form one sequence. A CTX token at each bar of the
target carries the context vector, a summary of the context. Only the
adapters, the projection and the new token embeddings
train.}]{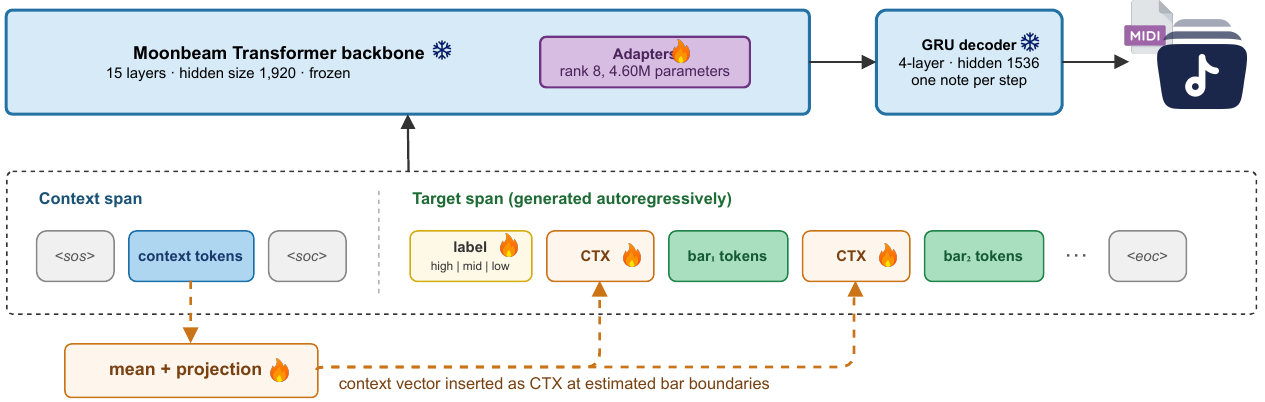}
\caption{A label and a context vector condition a frozen
pretrained model. Context, label and target form one sequence. A CTX
token at each bar of the target carries the context vector, a summary of
the context. Only the adapters, the projection and the new token
embeddings train.}
\end{figure}

\begingroup\singlespacing\small
\renewcommand{\arraystretch}{1.2}\hyphenpenalty=10000
\begin{longtable}[]{@{}
  >{\raggedright\arraybackslash}p{(\linewidth - 2\tabcolsep) * \real{0.1923}}
  >{\raggedright\arraybackslash}p{(\linewidth - 2\tabcolsep) * \real{0.8077}}@{}}
\multicolumn{2}{@{}p{\linewidth}@{}}{\normalsize \textbf{Table S2. Details Needed to Reproduce the Results}}\\[6pt]
\toprule\noalign{}
\begin{minipage}[b]{\linewidth}\raggedright
Item
\end{minipage} & \begin{minipage}[b]{\linewidth}\raggedright
Value
\end{minipage} \\
\midrule\noalign{}
\endfirsthead
\toprule\noalign{}
\begin{minipage}[b]{\linewidth}\raggedright
Item
\end{minipage} & \begin{minipage}[b]{\linewidth}\raggedright
Value
\end{minipage} \\
\midrule\noalign{}
\endhead
\bottomrule\noalign{}
\endlastfoot
Scores and data & NinSheetMusic MusicXML, converted to ABC for
annotation and to note arrays for the model; third-party transcriptions,
not redistributed. The annotations, pair lists, listening-test ratings
and code are public (see Code and Data Availability) \\
Annotation & Gemini 3.1 Pro, prompt version 1, thinking level high,
default sampling, JSON output; the prompt and output schema are part of
the code release \\
Pair labels & Thresholds 0.60 and 0.78 applied: the tertiles of CS over
all 3,304 pairs, test pieces included (0.6033 and 0.7779), rounded; a
pair at or above a threshold takes the higher label; fixed once, they
favour no label \\
Split & 236 / 49 / 49 pieces, stratified by section count, with 2,398 /
478 / 428 pairs; arrangements of one tune in one set \\
Sequence & One token per note, with its six fields; context and target
each truncated to 509 notes, so that both and five special tokens fit
1,024; one CTX token per target bar comes on top \\
Test contexts & Every main-body section of a test piece that is the
context of at least one test pair: 157 contexts from the 428 test pairs;
only main-body sections form pairs \\
Real gap & Means pooled over all real targets of a label in the test
contexts, each cut at its context's length like the outputs, since only
39 of the 157 contexts have both a high- and a low-similarity target; at
full length the real gaps are 0.238 on Melody and 0.362 on CS, and
within those 39 contexts 0.30 and 0.35 \\
Training & AdamW at a constant learning rate of \(3 \times 10^{-5}\),
batch size 2; seed 42; at most 200 epochs, early stopping with patience
20, lowest validation loss at epoch 11 (controls: 17, 14 and 18 for no
label, wrong label, unconditioned); one A100 GPU with 80 GB \\
Backbone and adapters & 15 layers, hidden size 1,920; adapters of rank 8
and scale 2, without dropout, on the query and value projections of all
15 attention layers; backbone and note decoder frozen \\
Trainable parameters & 4.60 million (parts rounded): adapters 0.81
million, context-vector projection 3.69 million, new token embeddings
0.11 million \\
Context vector & Mean of the backbone's last hidden layer over the
context, adapters active, then a learned linear projection \\
Generation & Draws numbered 0 to 5. Draw 0, the first, uses seed 42 as
one stream over contexts and labels; draw \(d \geq 1\) seeds each output
with \(42 + 100000 d + 10 i + j\), for context index \(i\) and label
index \(j\). The controls seed every draw, the first included, by the
same formula \\
Post-processing & Every output cut at the context's length; notes
emitted with a one-tick duration extended to the next onset on the same
MIDI channel \\
Bars & Bar length estimated as the context's duration over its number of
bars \\
Note budget & 1.2 times the mean length, in notes, of the context's test
targets; the same for every label \\
\end{longtable}
\par\endgroup

\FloatBarrier
\subsection{Label Comparisons}\label{supp-label-comparisons}

Table S3 extends Figure 2 to all 12 measures and gives the mean gap of
every comparison. High-similarity outputs lie closer to the context than
low-similarity outputs on all 12, with 10 significant after the
correction over all 12 measures. The main text corrects over its five.
They also exceed mid-similarity outputs on all 12, all after correction,
while mid-similarity outputs exceed low-similarity outputs on none.

\begingroup\singlespacing\small
\renewcommand{\arraystretch}{1.2}\hyphenpenalty=10000
\begin{longtable}[]{@{}
  >{\raggedright\arraybackslash}p{(\linewidth - 10\tabcolsep) * \real{0.1146}}
  >{\raggedright\arraybackslash}p{(\linewidth - 10\tabcolsep) * \real{0.1562}}
  >{\raggedleft\arraybackslash}p{(\linewidth - 10\tabcolsep) * \real{0.0833}}
  >{\raggedleft\arraybackslash}p{(\linewidth - 10\tabcolsep) * \real{0.2708}}
  >{\raggedleft\arraybackslash}p{(\linewidth - 10\tabcolsep) * \real{0.0833}}
  >{\raggedleft\arraybackslash}p{(\linewidth - 10\tabcolsep) * \real{0.2917}}@{}}
\multicolumn{6}{@{}p{\linewidth}@{}}{\normalsize \textbf{Table S3. All Twelve Measures Place High-Similarity Outputs
Closest to the Context}}\\[6pt]
\toprule\noalign{}
\begin{minipage}[b]{\linewidth}\raggedright
Measure
\end{minipage} & \begin{minipage}[b]{\linewidth}\raggedright
Comparison
\end{minipage} & \begin{minipage}[b]{\linewidth}\raggedleft
Share
\end{minipage} & \begin{minipage}[b]{\linewidth}\raggedleft
Mean Gap (95\% Interval)
\end{minipage} & \begin{minipage}[b]{\linewidth}\raggedleft
Holm p
\end{minipage} & \begin{minipage}[b]{\linewidth}\raggedleft
After the Filter: Mean Gap; Holm p
\end{minipage} \\
\midrule\noalign{}
\endfirsthead
\toprule\noalign{}
\begin{minipage}[b]{\linewidth}\raggedright
Measure
\end{minipage} & \begin{minipage}[b]{\linewidth}\raggedright
Comparison
\end{minipage} & \begin{minipage}[b]{\linewidth}\raggedleft
Share
\end{minipage} & \begin{minipage}[b]{\linewidth}\raggedleft
Mean Gap (95\% Interval)
\end{minipage} & \begin{minipage}[b]{\linewidth}\raggedleft
Holm p
\end{minipage} & \begin{minipage}[b]{\linewidth}\raggedleft
After the Filter: Mean Gap; Holm p
\end{minipage} \\
\midrule\noalign{}
\endhead
\bottomrule\noalign{}
\multicolumn{6}{@{}p{\linewidth}@{}}{\emph{Share of the 157 test contexts, from 49 pieces, in which the first
label's output, averaged over six draws, lies closer to the context. The
gap is in the measure's own units. Intervals and one-sided p come from
2,000 resamples of whole pieces, with the Holm correction over the 12
measures, per comparison; with 2,000 resamples its smallest value is
0.006, or 0.003 over five measures, as in Table 3. Contexts without a
gap are left out of the share. The last column repeats the gap and its
Holm p on the 149 contexts left by the stimulus filter; over the five
main-text measures, the filtered Holm p run from 0.003 to 0.026.}}\\
\endlastfoot
Melody & high \textgreater{} low & 62.7\% & +0.0080 (+0.0045 to +0.0121)
& 0.006 & +0.0094 (+0.0050 to +0.0142); 0.006 \\
& high \textgreater{} mid & 70.4\% & +0.0119 (+0.0081 to +0.0163) &
0.006 & +0.0132 (+0.0086 to +0.0184); 0.006 \\
& mid \textgreater{} low & 45.6\% & -0.0038 (-0.0064 to -0.0012) & 1.000
& -0.0038 (-0.0076 to +0.0000); 1.000 \\
Chroma & high \textgreater{} low & 55.4\% & +0.0137 (+0.0048 to +0.0227)
& 0.006 & +0.0121 (+0.0015 to +0.0228); 0.091 \\
& high \textgreater{} mid & 66.9\% & +0.0231 (+0.0157 to +0.0301) &
0.006 & +0.0167 (+0.0063 to +0.0271); 0.007 \\
& mid \textgreater{} low & 39.5\% & -0.0094 (-0.0175 to -0.0009) & 1.000
& -0.0046 (-0.0162 to +0.0083); 1.000 \\
Rhythm & high \textgreater{} low & 62.7\% & +0.0050 (+0.0019 to +0.0082)
& 0.009 & +0.0051 (+0.0018 to +0.0088); 0.025 \\
& high \textgreater{} mid & 70.6\% & +0.0066 (+0.0038 to +0.0094) &
0.006 & +0.0090 (+0.0039 to +0.0148); 0.006 \\
& mid \textgreater{} low & 47.1\% & -0.0016 (-0.0045 to +0.0014) & 1.000
& -0.0039 (-0.0095 to +0.0005); 1.000 \\
Duration & high \textgreater{} low & 61.1\% & +0.0154 (+0.0023 to
+0.0294) & 0.048 & +0.0150 (+0.0007 to +0.0309); 0.126 \\
& high \textgreater{} mid & 66.2\% & +0.0329 (+0.0198 to +0.0450) &
0.006 & +0.0402 (+0.0232 to +0.0559); 0.006 \\
& mid \textgreater{} low & 42.3\% & -0.0176 (-0.0308 to -0.0017) & 1.000
& -0.0252 (-0.0412 to -0.0079); 1.000 \\
CS (ours) & high \textgreater{} low & 58.6\% & +0.0132 (+0.0030 to
+0.0231) & 0.035 & +0.0212 (+0.0079 to +0.0365); 0.016 \\
& high \textgreater{} mid & 62.4\% & +0.0152 (+0.0063 to +0.0244) &
0.006 & +0.0185 (+0.0037 to +0.0318); 0.014 \\
& mid \textgreater{} low & 51.0\% & -0.0019 (-0.0099 to +0.0062) & 1.000
& +0.0027 (-0.0073 to +0.0133); 1.000 \\
Melody-Contour & high \textgreater{} low & 56.7\% & +0.0125 (-0.0004 to
+0.0248) & 0.060 & +0.0067 (-0.0043 to +0.0191); 0.382 \\
& high \textgreater{} mid & 55.6\% & +0.0266 (+0.0119 to +0.0406) &
0.006 & +0.0166 (-0.0006 to +0.0341); 0.029 \\
& mid \textgreater{} low & 41.9\% & -0.0141 (-0.0265 to -0.0029) & 1.000
& -0.0099 (-0.0254 to +0.0049); 1.000 \\
Melody-Intervals & high \textgreater{} low & 61.8\% & +0.0311 (+0.0124
to +0.0494) & 0.009 & +0.0176 (-0.0026 to +0.0386); 0.174 \\
& high \textgreater{} mid & 68.8\% & +0.0569 (+0.0410 to +0.0744) &
0.006 & +0.0404 (+0.0176 to +0.0616); 0.007 \\
& mid \textgreater{} low & 36.9\% & -0.0259 (-0.0424 to -0.0095) & 1.000
& -0.0229 (-0.0412 to -0.0051); 1.000 \\
Melody-Alignment & high \textgreater{} low & 59.9\% & +0.0064 (+0.0012
to +0.0117) & 0.035 & +0.0063 (+0.0002 to +0.0135); 0.126 \\
& high \textgreater{} mid & 56.1\% & +0.0112 (+0.0059 to +0.0169) &
0.006 & +0.0119 (+0.0058 to +0.0189); 0.006 \\
& mid \textgreater{} low & 49.0\% & -0.0048 (-0.0098 to -0.0003) & 1.000
& -0.0056 (-0.0113 to +0.0001); 1.000 \\
Rhythm-Overlap & high \textgreater{} low & 65.1\% & +0.0348 (+0.0175 to
+0.0530) & 0.006 & +0.0287 (+0.0089 to +0.0496); 0.025 \\
& high \textgreater{} mid & 66.2\% & +0.0387 (+0.0234 to +0.0535) &
0.006 & +0.0354 (+0.0144 to +0.0578); 0.006 \\
& mid \textgreater{} low & 51.3\% & -0.0039 (-0.0192 to +0.0116) & 1.000
& -0.0067 (-0.0268 to +0.0124); 1.000 \\
Rhythm-Grid & high \textgreater{} low & 61.8\% & +0.0065 (+0.0006 to
+0.0126) & 0.048 & +0.0039 (-0.0026 to +0.0101); 0.382 \\
& high \textgreater{} mid & 65.4\% & +0.0100 (+0.0051 to +0.0157) &
0.006 & +0.0099 (+0.0033 to +0.0173); 0.006 \\
& mid \textgreater{} low & 44.6\% & -0.0034 (-0.0089 to +0.0019) & 1.000
& -0.0060 (-0.0141 to +0.0013); 1.000 \\
Register & high \textgreater{} low & 53.5\% & +0.0021 (-0.0011 to
+0.0052) & 0.109 & +0.0054 (+0.0009 to +0.0099); 0.072 \\
& high \textgreater{} mid & 68.8\% & +0.0078 (+0.0041 to +0.0112) &
0.006 & +0.0071 (+0.0026 to +0.0112); 0.007 \\
& mid \textgreater{} low & 41.4\% & -0.0057 (-0.0096 to -0.0020) & 1.000
& -0.0016 (-0.0058 to +0.0026); 1.000 \\
Density & high \textgreater{} low & 57.3\% & +0.0183 (+0.0043 to
+0.0329) & 0.035 & +0.0103 (-0.0092 to +0.0289); 0.382 \\
& high \textgreater{} mid & 65.6\% & +0.0337 (+0.0224 to +0.0445) &
0.006 & +0.0309 (+0.0142 to +0.0485); 0.006 \\
& mid \textgreater{} low & 38.2\% & -0.0155 (-0.0289 to -0.0021) & 1.000
& -0.0206 (-0.0372 to -0.0031); 1.000 \\
\end{longtable}
\par\endgroup

\FloatBarrier
\subsection{Further Checks}\label{supp-further-checks}

These checks support the main text's account of the middle label and of
selection. The middle label produced the most unstable outputs; after
the stimulus filter, only Duration keeps its deficit against the low
label. The pairs themselves are ordered correctly: over all pairs,
mid-similarity targets are closer to their context than low-similarity
targets on all four main measures. Over all 334 pieces, among 290
contexts with both kinds of target, the mid-similarity target is closer
in 71.9\% on Melody (95\% interval 63.6--79.2\%). This check is
exploratory. Over all 3,304 pairs, low- and mid-similarity pairs share
one peak of CS, with means 0.556 and 0.653; high-similarity pairs lie
apart at 0.908.

For selection, we kept the one of twelve no-label draws with the highest
CS. It was compared with the expected single high-similarity draw, the
mean of six. On Melody, the gap is +0.0008, but its interval, -0.0066 to
+0.0082, is as wide as the label effect. On Chroma, selection came
closer by 0.020 (0.003 to 0.036), as expected, since CS-Chroma carries
four ninths of CS. This exploratory check leaves out CS, which did the
selecting. On Rhythm the gap is -0.0003 (-0.0058 to +0.0050) and on
Duration +0.0156 (-0.0100 to +0.0414), so the comparison separates
neither.

Restatement carries much of the Melody gap. We left out every output
whose top voice restates a run of eight or more notes of its context's
melody. This removes 116 high- and 67 low-similarity outputs of 929
each, and 132 of 1,880 no-label outputs. The high-minus-low gap on
Melody falls from 0.0080 to 0.0027, with an interval reaching zero (Holm
p = 0.083). Chroma, 0.0097, and Rhythm, 0.0045, stay significant over
the five measures; Duration and CS do not. Against no label, the
high-similarity label keeps its advantage on Melody, 0.0059 (0.0035 to
0.0088), and on Chroma, Rhythm and CS.

\FloatBarrier
\subsection{Stimulus Filter}\label{supp-stimulus-filter}

An output is unstable if it spans less than 55\% of the context or uses
fewer than 15\% of the context's distinct pitches. It is also unstable
if its melody has three pitches or fewer while the context's melody has
eight or more. The first two thresholds were set while the listening
stimuli were prepared. The third was added on 8 October 2026, after
listening to the drawn clips and before any rating was collected. Five
drawn listening contexts failed it and were replaced by the next five in
the same random order.

Of the 2,787 outputs in the 929 valid draws, 449 were unstable. Of
these, 126, 168 and 155 fell under the high, middle and low labels.
Output length does not favour the high-similarity label: after
truncation, outputs average 123, 128 and 129 notes under the high,
middle and low labels. Dropping every draw with an unstable output
removes 345 of 929 draws and leaves 149 contexts. On them, every
high-minus-low gap stays positive. All five main-text measures survive
the correction over five; over all 12, four do, CS among them. The
failure rates differ by label, so a second filter checks only the two
outputs that a comparison uses. It keeps 153 contexts; every gap stays
positive. Over five, Melody, Rhythm and CS survive the correction; over
12, five measures do.

\FloatBarrier
\subsection{Control Models and Context
Specificity}\label{supp-control-models-and-context-specificity}

The controls were trained with SimLlama's settings. They seed every
output by the formula of Table S2, the first draw included, where
SimLlama's first draw uses one stream. At least 99\% of their draws pass
the ten-note rule. Over all 2,826 outputs for the 157 contexts, the
ten-note failures included, the stimulus filter fails 471 of SimLlama's,
or 16.7\%. In valid draws it fails 449 of 2,787, or 16.1\%. It fails
15.7\% of the wrong-label control's, 14.1\% of the no-label control's
and 30.6\% of the unconditioned control's. The label-only model, trained
without the context vector at adapter rank 32 and drawn once, fails
39.7\%; it is not comparable in rank or draws. It gives outputs of ten
notes or more for 133 contexts. On them, it lies further from the
context than the unconditioned control on the four main measures and CS.

Table S4 repeats Table 3 on the contexts left by the stimulus filter.
After the filter, the high-similarity label has an advantage over no
label on all five measures, Duration included. The middle label's push
away survives on Duration only, and the context vector's gain on Chroma
only. Table S5 gives the test of context specificity for the four main
measures and CS. It scores each output against its own context and
against three contexts drawn at random from other test pieces.

\begin{table}[htbp]
\centering
\begin{minipage}{\linewidth}\small
\textbf{Table S4. Table 3 after the Stimulus Filter}
\par\medskip
\centering
\renewcommand{\arraystretch}{1.2}
\hyphenpenalty=10000
\begin{tabular}{@{}
  >{\raggedright\arraybackslash}p{(\linewidth - 10\tabcolsep) * \real{0.1220}}
  >{\raggedleft\arraybackslash}p{(\linewidth - 10\tabcolsep) * \real{0.1707}}
  >{\raggedleft\arraybackslash}p{(\linewidth - 10\tabcolsep) * \real{0.1707}}
  >{\raggedleft\arraybackslash}p{(\linewidth - 10\tabcolsep) * \real{0.1707}}
  >{\raggedleft\arraybackslash}p{(\linewidth - 10\tabcolsep) * \real{0.1707}}
  >{\raggedleft\arraybackslash}p{(\linewidth - 10\tabcolsep) * \real{0.1951}}@{}}
\toprule\noalign{}
\begin{minipage}[b]{\linewidth}\raggedright
Measure
\end{minipage} & \begin{minipage}[b]{\linewidth}\raggedleft
High \textgreater{} No Label
\end{minipage} & \begin{minipage}[b]{\linewidth}\raggedleft
No Label \textgreater{} Low
\end{minipage} & \begin{minipage}[b]{\linewidth}\raggedleft
No Label \textgreater{} Mid
\end{minipage} & \begin{minipage}[b]{\linewidth}\raggedleft
Wrong Label: High \textgreater{} Low
\end{minipage} & \begin{minipage}[b]{\linewidth}\raggedleft
No Label \textgreater{} Unconditioned
\end{minipage} \\
\midrule\noalign{}
Melody & \textbf{62\% (0.003)} & 51\% (1.000) & 56\% (1.000) &
\textbf{40\% (0.040)} & 50\% (0.948) \\
Chroma & \textbf{56\% (0.020)} & 50\% (1.000) & 55\% (1.000) & 47\%
(1.000) & \textbf{62\% (0.003)} \\
Rhythm & \textbf{68\% (0.003)} & 45\% (1.000) & 47\% (1.000) & 52\%
(1.000) & 55\% (0.248) \\
Duration & \textbf{57\% (0.020)} & 50\% (1.000) & \textbf{64\% (0.005)}
& 50\% (1.000) & 53\% (0.339) \\
CS (ours) & \textbf{62\% (0.003)} & 52\% (1.000) & 54\% (1.000) & 49\%
(1.000) & 52\% (0.948) \\
\bottomrule
\end{tabular}
\par\medskip
\noindent\emph{As Table 3, on the 149 contexts left by the stimulus filter (148
for the wrong-label column). Bold: Holm-corrected p \textless{} 0.05.}
\end{minipage}
\end{table}

\begin{table}[htbp]
\centering
\begin{minipage}{\linewidth}\small
\textbf{Table S5. On Melody, Chroma, Rhythm and CS, the Label Effect
Belongs to the Output's Own Context}
\par\medskip
\centering
\renewcommand{\arraystretch}{1.2}
\hyphenpenalty=10000
\begin{tabular}{@{}
  >{\raggedright\arraybackslash}p{(\linewidth - 8\tabcolsep) * \real{0.1579}}
  >{\raggedleft\arraybackslash}p{(\linewidth - 8\tabcolsep) * \real{0.2105}}
  >{\raggedleft\arraybackslash}p{(\linewidth - 8\tabcolsep) * \real{0.2105}}
  >{\raggedleft\arraybackslash}p{(\linewidth - 8\tabcolsep) * \real{0.2105}}
  >{\raggedleft\arraybackslash}p{(\linewidth - 8\tabcolsep) * \real{0.2105}}@{}}
\toprule\noalign{}
\begin{minipage}[b]{\linewidth}\raggedright
Measure
\end{minipage} & \begin{minipage}[b]{\linewidth}\raggedleft
Gap, Own Context
\end{minipage} & \begin{minipage}[b]{\linewidth}\raggedleft
Gap, Foreign Contexts
\end{minipage} & \begin{minipage}[b]{\linewidth}\raggedleft
Difference (95\% Interval)
\end{minipage} & \begin{minipage}[b]{\linewidth}\raggedleft
Holm p
\end{minipage} \\
\midrule\noalign{}
Melody & 0.0080 & -0.0001 & 0.0081 (0.0041 to 0.0126) & 0.003 \\
Chroma & 0.0137 & 0.0057 & 0.0080 (0.0010 to 0.0149) & 0.030 \\
Rhythm & 0.0050 & 0.0012 & 0.0038 (0.0009 to 0.0069) & 0.030 \\
Duration & 0.0154 & 0.0051 & 0.0103 (-0.0036 to 0.0248) & 0.069 \\
CS (ours) & 0.0132 & -0.0003 & 0.0136 (0.0033 to 0.0236) & 0.030 \\
\bottomrule
\end{tabular}
\par\medskip
\noindent\emph{Gaps are high minus low, averaged over draws, then over the 157
contexts. Intervals and one-sided p come from 2,000 resamples of whole
pieces, with the Holm correction over these five measures.}
\end{minipage}
\end{table}

\FloatBarrier
\subsection{Listener Agreement}\label{supp-listener-agreement}

Table S6 extends Table 4 of the main text to all 12 measures, scored on
the 10 seconds that listeners heard. Five measures survive the Holm
correction. Melody, Melody-Intervals and Melody-Contour are also the
three largest effects; Register and Density lose their effect when
entered together with Melody. Without the four clips that restate their
context, Melody-Intervals, Density, Melody-Contour and Melody lead with
3.6 to 4.1 points per standard deviation. These four survive the
correction. Melody scores zero for 28 clips, which share no
three-interval pattern with their context. Without them, its rise is 7.0
points. A by-listener random slope for each measure changes no estimate
by more than 0.1 points, and the same five measures survive the
correction.

Entered alone with CS, Melody rises 6.3 points (4.0 to 8.6). It exceeds
CS by 5.0 (1.2 to 8.7), and without the four restating clips by 3.1
(-1.5 to 7.6). The per-trial Spearman of Figure 4 is 0.50 (p = 0.009)
without the three trials that hold a restating clip.

\begingroup\singlespacing\small
\renewcommand{\arraystretch}{1.2}\hyphenpenalty=10000
\begin{longtable}[]{@{}lrrrrr@{}}
\multicolumn{6}{@{}p{\linewidth}@{}}{\normalsize \textbf{Table S6. The Three Strongest Predictors of Listener Ratings All
Follow the Melody}}\\[6pt]
\toprule\noalign{}
Measure & Rise per SD & 95\% Interval & p & Holm p & r over 60 Clips \\
\midrule\noalign{}
\endfirsthead
\toprule\noalign{}
Measure & Rise per SD & 95\% Interval & p & Holm p & r over 60 Clips \\
\midrule\noalign{}
\endhead
\bottomrule\noalign{}
\multicolumn{6}{@{}p{\linewidth}@{}}{\emph{20 listeners, 60 clips. Mixed model with a slot term and random
intercepts for listener, context and clip; Wald 95\% intervals. The last
column is the correlation of each measure with the mean rating of each
clip. Eight variants of a revised label measure, not in this paper, were
also scored; none exceeds Melody, which also survives a Holm correction
over all 20.}}\\
\endlastfoot
Melody & +6.7 & 4.60 to 8.86 & \textless{} 0.001 & \textless{} 0.001 &
0.62 \\
Chroma & +2.9 & 0.20 to 5.51 & 0.035 & 0.211 & 0.27 \\
Rhythm & +2.7 & -0.04 to 5.42 & 0.054 & 0.224 & 0.23 \\
Duration & +1.5 & -1.25 to 4.30 & 0.281 & 0.442 & 0.14 \\
CS (ours) & +3.4 & 0.87 to 6.01 & 0.009 & 0.060 & 0.30 \\
Melody-Contour & +5.0 & 2.69 to 7.40 & \textless{} 0.001 & \textless{}
0.001 & 0.49 \\
Melody-Intervals & +5.6 & 3.33 to 7.94 & \textless{} 0.001 & \textless{}
0.001 & 0.54 \\
Melody-Alignment & +2.9 & 0.07 to 5.71 & 0.045 & 0.224 & 0.25 \\
Rhythm-Overlap & +1.7 & -1.02 to 4.44 & 0.221 & 0.442 & 0.17 \\
Rhythm-Grid & +2.4 & -0.32 to 5.16 & 0.083 & 0.250 & 0.22 \\
Register & +3.7 & 1.26 to 6.20 & 0.003 & 0.025 & 0.36 \\
Density & +4.1 & 1.47 to 6.78 & 0.002 & 0.021 & 0.33 \\
\end{longtable}
\par\endgroup

\FloatBarrier
\subsection{Listening Details}\label{supp-listening-details}

Moonbeam emits an instrument per note; 13.3\% of the outputs hold a
non-piano note, 2.9\% of all notes. Every instrument was set to piano
before rendering, so timbre could not tell the two outputs apart. Notes
that start before 10 seconds may ring for one more second, so the clips
last 9.8 to 11.5 seconds. The measures score the same notes. The
attention checks were trials 11 and 21, in which the real section shares
the context's section letter. All listeners received the trials in one
order, shuffled once with a recorded seed, with the same A/B assignment.
Within a trial, they could play and replay the three clips in any order.
Valid sessions lasted 39 to 56 minutes. Of the 20 listeners, 2 reported
no formal musical training, 11 up to three years, 6 four to ten and 1
more than ten.

Over listeners, 17 of 20 favoured the high-similarity output, by a mean
of 2.9 points (one-sided Wilcoxon test, p \textless{} 0.001). This test
treats the 30 trials as fixed. Over trials, 18 of 30 did (p = 0.159).
The mean gap was +1.8 points over the first half of the trials and +4.0
over the second. The halves did not differ significantly (p = 0.202);
they also hold different contexts. With the six excluded listeners
restored, 20 of 26 favoured the high-similarity output. The label effect
is then +2.2 points (-1.8 to +6.2, p = 0.27). Melody still predicts the
ratings best, by 5.6 points per standard deviation (3.9 to 7.3, Holm p
\textless{} 0.001). The minimum detectable label effect, at 80\% power
and two-sided 0.05, is 2.80 times its standard error of 2.47, or 6.9
points.

\end{document}